\documentclass[12pt]{article}

\usepackage{fancyhdr}
\usepackage{isomath}
\usepackage{amsmath}
\usepackage{amsbsy}
\usepackage{amssymb}
\usepackage{amscd}
\usepackage{amsfonts}
\usepackage{graphics}
\usepackage{verbatim}
\usepackage{xspace}
\usepackage{euscript}
\usepackage{alltt}
\usepackage{boxedminipage}
\usepackage{float}
\usepackage{color}
\usepackage{color,soul}
\usepackage{t1enc}
\usepackage{times,exscale}
\usepackage{graphicx,calc}
\usepackage{mdwlist}
\usepackage{stmaryrd}
\usepackage{units}
\usepackage{setspace}
\usepackage{pdfpages}
\usepackage{hyperref}
\usepackage[usenames,dvipsnames]{xcolor}
\usepackage{cite}
\usepackage[normalem]{ulem}

\usepackage[english]{babel}
\usepackage[utf8]{inputenc}
\usepackage{algorithm}
\usepackage[noend]{algpseudocode}
\usepackage{gensymb}
\usepackage{CJKutf8}

\graphicspath{ {imagedir/} }

\usepackage[margin=20mm]{geometry}
\numberwithin{equation}{section}

\title{A 3D phase field dislocation dynamics model for body-centered cubic crystals}
\author{
	Xiaoyao Peng$^{\P}$, 
	Nithin Mathew$^{\dagger}$, 
	Irene J. Beyerlein$^{\ddagger}$, 
	Kaushik Dayal$^{\S\P}$, 
	Abigail Hunter\footnote{Email: \url{ahunter@lanl.gov}}
	\\ 
	\small{$^{\P}$Department of Civil and Environmental Engineering, Carnegie Mellon University}
	\\
	\small{$^{\dagger}$Theoretical Division, T-1/Center for Nonlinear Studies, Los Alamos National Laboratory}
	\\
	\small{$^{\ddagger}$University of California, Santa Barbara}
	\\
	\small{$^{\S}$Center for Nonlinear Analysis, Carnegie Mellon University}
	\\
	\small{$^{\S}$Department of Materials Science and Engineering, Carnegie Mellon University}
	\\
	\small{$^{\ast}$X Computational Physics Division, Los Alamos National Laboratory}
}
\date{\today}

 \usepackage{epsfig}

\usepackage{amssymb}
\usepackage{amsmath}

\usepackage{fancyhdr}
\usepackage{isomath}
\usepackage{amsbsy}
\usepackage{amscd}
\usepackage{amsfonts}
\usepackage{verbatim}
\usepackage{xspace}
\usepackage{euscript}
\usepackage{alltt}
\usepackage{boxedminipage}
\usepackage{float}
\usepackage{color}
\usepackage{t1enc}
\usepackage{times,exscale}
\usepackage{mdwlist}
\usepackage{stmaryrd}
\usepackage{units}
\usepackage{setspace}
\usepackage{pdfpages}
\usepackage{hyperref}
\usepackage[usenames,dvipsnames]{xcolor}
\usepackage[normalem]{ulem}

\usepackage[english]{babel}
\usepackage[utf8]{inputenc}
\usepackage{algorithm}
\usepackage[noend]{algpseudocode}
\usepackage{gensymb}
\usepackage{CJKutf8}

 \usepackage{color,soul}
 \soulregister\citep7
 \soulregister\ref7
 \usepackage{multirow}
 \usepackage{subfigure}
 \usepackage{graphicx}

\newcommand{\bfb}{{\mathbold b}}
\newcommand{\bfx}{{\mathbold x}}
\newcommand{\bfs}{{\mathbold s}}
\newcommand{\bfk}{{\mathbold k}}
\newcommand{\bfm}{{\mathbold m}}
\newcommand{\bfn}{{\mathbold n}}
\newcommand{\bft}{{\mathbold t}}
\newcommand{\bfC}{{\mathbold C}}
\newcommand{\divergence}{\mathop{\rm div}\nolimits}
\newcommand{\bfsigma}{\mathbold {\sigma}}
\newcommand{\bfepsilon}{\mathbold {\epsilon}}
\def\Xint#1{\mathchoice
   {\XXint\displaystyle\textstyle{#1}}%
   {\XXint\textstyle\scriptstyle{#1}}%
   {\XXint\scriptstyle\scriptscriptstyle{#1}}%
   {\XXint\scriptscriptstyle\scriptscriptstyle{#1}}%
   \!\int}
\def\XXint#1#2#3{{\setbox0=\hbox{$#1{#2#3}{\int}$}
     \vcenter{\hbox{$#2#3$}}\kern-.5\wd0}}

\def\dashint{\Xint-}

\begin{document}

\pagestyle{fancyplain}
\lhead{\fancyplain{\scriptsize A 3D phase field dislocation dynamics model for body-centered cubic crystals \hfill X. Peng, N. Mathew, I. J. Beyerlein, K. Dayal, A. Hunter \\ To appear in {\em Computational Materials Science}; \url{doi.org/10.1016/j.commatsci.2019.109217}
	}
	{\scriptsize A 3D phase field dislocation dynamics model for body-centered cubic crystals \hfill X. Peng, N. Mathew, I. J. Beyerlein, K. Dayal, A. Hunter \\ To appear in {\em Computational Materials Science}; \url{doi.org/10.1016/j.commatsci.2019.109217}}
}
\rhead{\fancyplain{\scriptsize \quad} {\scriptsize \quad}}

\maketitle

\begin{abstract}

In this work, we present a 3D Phase Field Dislocation Dynamics (PFDD) model for body-centered cubic (BCC) metals. The model formulation is extended to account for the dependence of the Peierls barrier on the line-character of the dislocation.  Simulations of the expansion of a dislocation loop belonging to the $\left\{110\right\} \left<111\right>$ slip system are presented with direct comparison to Molecular Statics (MS) simulations. The extended PFDD model is able to capture the salient features of dislocation loop expansion predicted by MS simulations. The model is also applied to simulate the motion of a straight screw dislocation through kink-pair motion.
 
\end{abstract}



\section{Introduction}
\label{sec:introduction}

Dislocations are carriers of plasticity in metals \cite{Hull_Bacon:2001,Hirth:1968,Kocks:2003}.  During straining, individual crystals that are up to microns across will contain a collection of dislocations gliding on particular crystallographic slip planes in particular slip directions.  They move and are stored heterogeneously throughout the crystal.  Dislocations or groups of dislocations have been represented in various ways, atomically including their core to discretely as a line to statistically as a dislocation density to finally crystallographically as slip.  Models that represent dislocations discretely, including their Burgers vector and line orientation, have thus far been successful in capturing heterogeneity in slip.  

Atomistic approaches, including Molecular dynamics (MD), Molecular Statics (MS), and Density Functional Theory (DFT), are one such example.  These methods can model the movement of a few dislocations in small volumes (high dislocation densities) and account for the atomic-scale effects of the dislocation core during motion along with interactions and reactions with other dislocations and boundaries in the system \cite{Lesar:2013, Zepeda-Ruiz2017, GWang:2004, Weinberger:2013, Lim:2018}.  The chief disadvantage of atomic-scale techniques is that they are limited in the material sizes and time scales they can access.  

Moving up in length scale, mesoscale dislocation-dynamics (DD) codes have been emerging over the recent past in an effort to overcome the short time and length scales that limit atomistic methods. These models account for each dislocation as a line in a 3D continuum that interacts with other linear dislocations via elastic fields.  The motion and interaction of the individual dislocation lines in a system is modeled over time and under various applied loading conditions.  These approaches do not resolve the dislocation lines to the atomic level, hence features such as the dislocation cores cannot be explicitly captured.  

One such DD model, classified as discrete DD techniques (DDD), was developed in the late 90s \cite{Ghoniem:2000, Devincre:1997, Rhee:1998, Weygand:2002}. Since then DDD models have become a well-developed approach for modeling a group to several interacting dislocation lines in 3D space.  They have been successful in addressing the evolution of many interacting dislocations \cite{Zbib:2011, Cai2:2004, Kubin:1998, ZWang:2008, Zhou:2011} and the formation of organized dislocation structures in strained crystals \cite{Madec2:2002,Kubin2:1992,Kubin:1993,Argaman:2001,ZWang:2009,Dmitrieva:2010}. They have also considered a variety of crystal structures including face-centered cubic (FCC), body-centered cubic (BCC) metals and hexagonal close-packed (HCP) metals \cite{Zbib:2000, Arsenlis:2007, Bertin:2014, Monnet:2004, Lesar:2013, BulatovCai:2006}.  They were originally developed for bulk crystals but in the past decade have been further developed to account for the barrier effects of interfaces and grain boundaries, twin boundaries, precipitates, and thin films \cite{ElAwady:2016,Hussein:2017,Gao:2015}.

In DDD models, the dislocation line is discretized into a finite number of segments, and at each point along the line the equation of motion is solved. Some elementary processes, such as dissociation, climb, non-Schmid effects, or cross slip can be modeled with the help of rules, typically motivated by atomistic simulations, including MS, MD, or DFT \cite{ElAwady:2008, ZWang:2011, JWang:2014, Shehadeh:2007, ZWang:2011}.  For instance, the model by Shehadeh \textit{et al.} \cite{Shehadeh:2007} incorporates fault energies into the DDD model to simulate slip transmission across an interface, including the possibility of core spreading in the interface.  A number of DDD models have incorporated the effects of cross slip \cite{ZWang2:2007, Kubin:1992, ZWang:2008}, finding that they lead to formation of slip bands or cellular structures. A recent set of works have applied to examine the effects of precipitates in superalloys \cite{Hussein:2017, Yang:2013, Huang:2012, Gao:2015}.

Another type of mesoscale mechanics technique for simulating dislocations that has emerged is phase field dislocation dynamics (PFDD).  Phase field models have been traditionally used to study phase transformations \cite{Steinbach:2009}.  In the last few decades, they been extended to model dislocation behavior and interactions \cite{Wang2001, Koslowski2002}, giving rise to the name PFDD. PFDD adopts the original basis of phase field models in the sense that system dynamics are directly dependent on minimization of the total system energy.  However, in contrast to traditional phase field approaches, the order parameters in PFDD represent individual dislocations rather than different material phases.  More specifically, in PFDD, dislocations in each slip system, $\alpha$, are represented by scalar order parameters, $\zeta^{\alpha}(\bfx,t)$; one order parameter per slip plane. At any given point on a slip plane, this phase field parameter records the amount of slip that has occurred due to dislocation motion in units of the Burgers vector \cite{Koslowski2002}.  Transitions in the order parameter can represent the locations of dislocation lines. A master energy functional is comprised of the elastic strain energy, calculated from continuum linear elastic dislocation theory, and the dislocation lattice energy, which is often informed from atomic scale simulations.  The equilibrium positions of the dislocations in the crystal are determined by minimizing this energy functional at every time step. In this way, PFDD enables simulations with much larger crystal sizes and time scales (seconds and 10-100 nanometers), more comparable to those used in experiments, than atomic-scale simulations. 

The PFDD model was developed to simulate the evolution of dislocations in any material.  Despite this, most PFDD studies have focused on modeling dislocations in FCC materials, with a few modeling BCC dislocations \cite{Beyerlein2016,Mori:2011}. In FCC metals, PFDD studies have considered not only perfect dislocations but also partial dislocations \cite{Beyerlein2016, Hunter2:2014, Hunter:2011}, heterophase interfaces \cite{Zeng:2016, Hunter:2018}, and deformation twinning \cite{Hunter3:2014, Hunter:2015}.  In recent years, they have successfully been advanced to model grain boundary sliding \cite{Koslowski:2011}, grain boundary dislocation nucleation \cite{Hunter:2014, Hunter2:2014}, glide in high entropy alloys \cite{Zeng:2019}, texture effects in thin films \cite{Cao:2017}, formation and glide of partial dislocations in polycrystals \cite{Cao:2015}, the presence of void space \cite{Lei:2013}, and slip transfer across biphase boundaries \cite{Zeng:2016, Hunter:2018}, again, all for FCC metals.  Recent advancements have been made in predicting dislocation networks in boundaries of BCC metals \cite{Qiu:2019} and transformations to the HCP phase by sequential glide of FCC Shockley partials \cite{Louchez:2017}.

Dislocation motion in FCC materials is fundamentally simpler than that in other metals.  First, FCC materials deform predominantly via one slip mode. In comparison, it is well known that BCC metals slip along the close-packed direction, $\left<111\right>$, however the slip planes on which slip occurs is harder to definitively identify (for a detailed review on this topic see \cite{Weinberger2:2013,Christian:1983}). Experimental evidence has observed slip on $\left\{110\right\}$, $\left\{112\right\}$, and $\left\{123\right\}$ planes \cite{Weinberger2:2013,Lim:2019,Hirth:1968,Hsiung:2010,Bowen:1967,Shields:1975,Richter:1971}.  What still remains controversial is whether or not slip is really being activated on $\left\{112\right\}$ or $\left\{123\right\}$ planes, or if composite or aggregate slip on $\left\{110\right\}$ planes are producing a net $\left\{112\right\}$ (or $\left\{123\right\}$) slip that is then observed \cite{Weinberger2:2013,Lim:2019,Hirth:1968}.  To add further complexity, the activated slip mode can depend heavily on the loading conditions (i.e, strain rate, temperature, orientation), which is largely a result of the unique, non-planar core structure of screw dislocations in BCC metals \cite{Weinberger2:2013, Christian:1983,Rodney:2017}.  For example, this core structure produces high lattice resistance causing the glide of screw dislocations to be thermally activated through kink nucleation mechanisms at low temperatures.  In addition, BCC metals are well-known for non-Schmid behavior, in particular showing a distinct tension-compression asymmetry under uniaxial loading, which is much larger than that seen in FCC metals \cite{Duesbery:1998,Vitek:2008}.  Deviation from the Schmid law is due to two distinct effects: the critical resolved shear stress can be influenced by any component of the applied stress tensor, and the critical resolved shear stress is not independent of slip system and sense of slip \cite{Duesbery:1998}.  The first of these is due to the non-planar core structure of screw dislocations in BCC metals responding to different components of the applied stress tensor than expected \cite{Vitek:2008,Groger:2019}.  The second aspect is the well-known twinning/anti-twinning slip asymmetry found in BCC metals \cite{Duesbery:1998,Dezerald:2016,ZWang:2011}.  Finally, the core structure of screw dislocations in BCC metals is different from the core structure of edge dislocations.   Hence, and also in contrast to FCC dislocations, the motion of dislocations in BCC metals depends on the line-character \cite{Urabe:1975, Low:1962, Byron:1967, Tang:1998}.  These distinctions induce changes in the types of active deformation mechanisms in the metal. For example, low temperature electron microscopy measurements in BCC crystals have shown the presence of long screw dislocations, indicating that non-screw dislocations are more mobile and require a lower resolved shear stress (RSS) for motion \cite{Hull_Bacon:2001}. Due to the high RSS required, screw dislocations in BCC metals move by nucleation and propagation of kink-pairs. A kink-pair is first nucleated along the screw-oriented dislocation dislocation line. The kink is bound by a pair of oppositely signed edge oriented components. Due to the low RSS required to propagate edge-oriented dislocations, the sides of the kink can move easily laterally along the dislocation line and the screw dislocation ultimately advances.  Thus, to accurately capture dislocation motion in BCC materials, it is important to account for the line-character and capture the dependence of RSS on the line-character. 

In this work, we present a 3D PFDD framework that includes dislocation character dependent behavior applicable to BCC metals. The article is structured as follows. First, the PFDD framework traditionally used for perfect dislocation motion in FCC materials is briefly presented in Section \ref{sec:pfdd}. Extensions to account for a line-character dependent Peierls barrier are presented next in Section \ref{ChapCoreEnergy}. Two cases are treated with the extended BCC PFDD model in Section \ref{sec:results}, dislocation loop expansion and the kink-pair motion of a screw dislocation, the former of which is presented with direct comparison to atomistic results.

\section{Methodology}

In this section, the PFDD formulation that has been traditionally used for perfect dislocations in FCC metals is first reviewed \cite{Beyerlein2016}.  Since BCC metals typically do not mediate plasticity through partial dislocation motion, the formulation for perfect dislocations is extended to account for differences in edge/screw dislocation motion that is common to BCC metals.  This extension incorporates a character dependence into the energy functional minimized in the PFDD formulation.   

\subsection{Phase Field Dislocation Dynamics (PFDD)}
\label{sec:pfdd}

As briefly described in the previous section, in the PFDD formulation, dislocations are described using scalar-valued order parameters $\zeta^{\alpha}(\bfx,t)$ defined on each active slip system $\alpha$.  For example, in an FCC material there can be up to 12 active order parameters, one parameter per slip system. Integer values of these order parameters represent perfect Burgers vector translations (or no translation at all with a zero value order parameter), where atomic bonds have already been broken and reformed and the crystal has undergone slip. A positive perfect dislocation on system corresponds to a positive integer jump in the order parameters and vice versa for a negatively signed dislocation. Order parameters with values between zero to one indicate the location of the dislocation core, where the atoms in the crystal structure are broken and not perfectly aligned. The total system energy $E$, can be expressed as a function of these order parameters.  To capture the motion of the dislocation network, the system is expected to evolve toward a minimum energy state. Minimization of the total energy is done using a time-dependent Ginzburg-Landau (TDGL) kinetic equation, which also allows for stress equilibrium, $ \divergence\bfsigma = \bf0 $, to be achieved:

\begin{equation}
\label{eq:TDGL}
    \frac{\partial\zeta^{\alpha}(\bfx,t)}{\partial t} = -L\frac{\partial E(\zeta)}{\partial \zeta^{\alpha}(\bfx,t)}
\end{equation}

\noindent with the assumption that the total strain $\bfepsilon(\bfx,t)=\bfepsilon^e(\bfx,t)+\bfepsilon^p(\bfx,t)$, where $\bfepsilon^e$ is the elastic strain, and $\bfepsilon^p$ is the plastic strain. The coefficient $L$ governs the rate at which equilibrium is achieved.  

We assume that plasticity is mediated by the motion and interaction of dislocations, and hence the plastic strain, $\bfepsilon^p$, is directly dependent on the active order parameters in a system \cite{Koslowski2002, Wang2001}, following

\begin{equation}
\label{eq:pstrain}
    \epsilon^p_{ij}(\bfx,t)=\frac{1}{2} \sum_{\alpha=1}^N b\zeta^{\alpha}(\bfx,t)\delta_\alpha(s_i^\alpha m_j^\alpha + s_j^\alpha m_i^\alpha),
\end{equation}

\noindent where $N$ is the number of active slip systems, $b$ is the magnitude of the Burgers vector $\bfb$, $\bfs^\alpha$ is the slip direction (normalized Burgers vector $\bfs\equiv\frac{\bfb}{b}$), $\bfm^\alpha$ is the slip plane normal $\alpha$, and $\delta_\alpha$ is a Dirac distribution supported on the slip plane of slip system $\alpha$.

The total system energy consists of two key energy terms \cite{Koslowski2002}:

\begin{equation}
\label{eq:TotalEnergy}
    E=E^{strain}+E^{lattice}.
\end{equation}

\noindent The first term, $E^{strain}$, is the strain energy, which accounts for elastic interactions between dislocations (e.g., attraction and repulsion), and interactions between the applied stress and the dislocations. It can be written as the sum of internal and external interaction terms:

\begin{equation}
\label{eq:Estrain}
    E^{strain}=E^{int}+E^{ext}=\frac{1}{2}\int C_{ijkl}\epsilon^e_{ij}(\bfx,t)\epsilon^e_{kl}(\bfx,t)d^3x - \int\sigma^{appl}_{ij}\epsilon^p_{ij}d^3x,
\end{equation}

\noindent where $\bfC$ is the elasticity tensor, and $\bfsigma^{appl}$ is the applied stress. $\bfepsilon^e$ can be solved in terms of $\bfepsilon^p$ through transformation into Fourier space, which provides the following expression of the internal strain energy \cite{Koslowski2002}:

\begin{equation}
    E^{int}(\zeta)=\frac{1}{2}\dashint \Hat{A}_{mnuv}(\bfk)\Hat{\epsilon}^p_{mn}(\bfk) \Hat{\epsilon}^{p*}_{uv}(\bfk) \frac{d^3k}{(2\pi)^3}
\label{eq:Eelast}
\end{equation}

\noindent where $\Hat{A}_{mnuv}(\bfk)=C_{mnuv}-C_{kluv}C_{ijmn}\Hat{G}_{ki}(\bfk)k_j k_l$, ($\string^$) denotes the Fourier transform, and ($\string*$) denotes the complex conjugate, $\bfk$ is the wave vector, $\mathbold{\Hat{G}}$ is the Fourier transform of the Green's tensor of linear elasticity \cite{mura1987general}, and the $\dashint$ denotes the Cauchy principal value of the integral.

The second term in Equation (\ref{eq:TotalEnergy}), $E^{lattice}$, accounts for the energy expended on breaking and reforming atomic bonds as dislocations move through the crystal lattice. Due to the periodicity of the crystal lattice, for perfect dislocations, $E^{lattice}$ is modeled with as a sinusoidal function \cite{Wang2001,Beyerlein2016}:

\begin{equation}
\label{eq:CoreEnergy}
    E^{lattice}(\zeta)=\sum_{\alpha=1}^N\int B\sin^2({n\pi\zeta^\alpha(\bfx,t)})\delta_\alpha d^3x
\end{equation}

\noindent where $B$ is the Peierls barrier, or the magnitude of the energy barrier to activate slip. This parameter can be informed many ways. For a wide range of BCC metals the Peierls potential for screw dislocations have been calculated using DFT \cite{Weinberger:2013, Lim:2018}.  It can also be calculated more quickly via MD simulations but with the caveat that accuracy depends on the reliability of the interatomic potential used \cite{Lesar:2013,Lim:2018,Zhang2:2011}.  More often, this term is informed using information from the material $\gamma$-surface, such as specific stacking fault energies or generalized stacking fault energy (GSFE) curves as calculated with atomistic methods \cite{Schoeck:2005, Vitek:1968}. Such energetic calculations do not involve the motion of a dislocation but rather shifts of two crystal halves with respect to one another across a crystallographic plane, and therefore do not correspond directly to the Peierls potential.  A relationship has been determined between the Peierls potential and the energy of the atomic bonds across a plane as a function of shifts in the atomic positions \cite{Joos:1997,Peierls:1940,Schoeck:2005}.  In BCC metals, the peak value of this energy function in the unstable stacking fault energy ($\gamma_U$), which can therefore be related to motion of the entire dislocation.

\subsection{Extension to BCC Metals}
\label{ChapCoreEnergy}

It has been shown that in BCC metals the Peierls stress for screw dislocations is one to two orders of magnitude greater than that for edge dislocations \cite{Louchet:1979, Vitek:1974}.  Consequently edge dislocations move faster through the crystal lattice than screw-type dislocations which, in turn, dominate the plastic response \cite{Hull_Bacon:2001, Hirth:1968, Kang2012}. To better model the plastic deformation in BCC metals, we extend the PFDD model to account for differences between edge and screw dislocation motion.

Specifically these extensions primarily modify $E^{lattice}$, so that character dependence of the Peierls energy barrier can be accounted for.  As mentioned above, the Peierls barrier itself can be estimated with atomistic modeling approaches.  However, to determine the Peierls potential for the all (or even many) dislocation characters is computationally costly.  In an effort to approximate the variation of the Peierls potential and also move up in length-scale, we utilize $\gamma_U$ in conjunction with a transition function.  The transition function captures the dependence of the Peierls energy barrier on screw/edge character of a dislocation.  Furthermore, this function describes how the energy barrier varies with character in general, and by incorporating such a term into $E^{lattice}$ dependence on dislocation line character is added into the PFDD model. Determining the line character of a general dislocation line requires calculation of the gradient of the order parameter. This is followed by, the development of the transition function which is a function of the character angle, $\theta$.

\subsubsection{Calculating the Dislocation Line Sense}

\begin{figure}[h]
\centering
\includegraphics[scale=1.0]{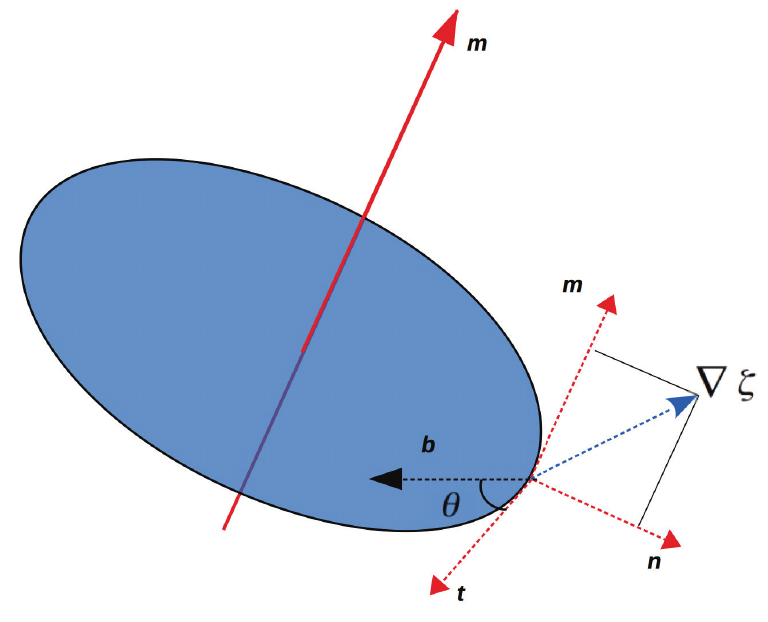}
\caption{\label{fig:grad-decomp} A dislocation loop (black curve surrounding the slipped region (denoted in blue)) with a Burgers vector $\bfb$. The gradient of the order parameter $\nabla\zeta^\alpha$ lies in the plane formed by dislocation line normal $\bfn$ and slip plane normal $\bfm$.}
\end{figure}

Figure \ref{fig:grad-decomp} presents a schematic of a general dislocation loop, with a local coordinate system comprised of orthogonal vectors: one is normal to the glide plane (also the slip plane normal $\bfm$ as defined in the global coordinate system), another is normal to the dislocation line $\bfn$, and a third that is the local tangent to the dislocation line $\bft$.  Within this system, the Burgers vector could be any vector within the $\bft$-$\bfn$ plane.  The tangent vector $\bft$ defines the line sense of the dislocation segment at that point along the loop.  The angle between this vector and the Burgers vector will define the character of that segment of dislocation, with an angle of $90^{\circ}$ indicating dislocations of pure edge type, and an angle of $0^{\circ}$ representing segments of pure screw type.

Within the PFDD model, such a loop would be modeled with a single active order parameter with a value of unity inside the loop and a value of zero outside the loop.  Hence, the gradient of the order parameter, $\nabla\zeta^\alpha$, will only be non-zero at points along the dislocation line, where the order parameter transitions between 0 and 1. In addition, the gradient of the order parameter lies in the $\bfn$-$\bfm$ plane, as shown in Figure \ref{fig:grad-decomp}, and defines the direction in which the order parameter has the most dramatic change.

In order to add character dependence to the total system energy the character angle, $\theta$, is needed for all points along all dislocation lines present in the simulation.  To achieve this we first calculate the gradient of all active order parameters across all computational points within a simulation using a central difference approach,

\begin{equation}
    \nabla\zeta^\alpha(\bfx,t) = \left(\frac{\zeta^\alpha(x_{i+1},t) - \zeta^\alpha(x_{i-1},t)}{2\Delta x}, \frac{\zeta^\alpha(y_{j+1},t) - \zeta^\alpha(y_{j-1},t)}{2\Delta y}, \frac{\zeta^\alpha(z_{k+1},t) - \zeta^\alpha(z_{k-1},t)}{2\Delta z}\right)
\end{equation}

\noindent where $(x_i, y_j, z_k)$ is any computational grid point, and $(\Delta x, \Delta y, \Delta z)$ are the grid spacings in the $(x,y,z)$ directions, respectively.  Taking the cross product of this gradient vector with the slip plane normal produces the vector tangent to the dislocation line, i.e., the line sense.  This unit tangent vector can be expressed as:

\begin{equation}
\label{eq:tang}
    \bft^\alpha(\bfx,t)=\frac{\nabla\zeta^\alpha(\bfx,t)\times\bfm^\alpha}{|\nabla\zeta^\alpha(\bfx,t)\times\bfm^\alpha|}.
\end{equation}

\noindent Finally, the character angle, $\theta^\alpha$, can be calculated by taking a dot product of the tangent vector and the slip direction (i.e., the normalized Burger vector):

\begin{equation}
\label{eq:theta}
    \theta^\alpha(\bfx,t)=\cos^{-1}{\bft^\alpha(\bfx,t)\cdot\bfs^\alpha}.
\end{equation}

\noindent Recall, $\theta^\alpha=90^{\circ}$ represents a point on the dislocation loop that is of pure edge type, and $\theta^\alpha=0^{\circ}$ indicates pure screw type. Other values between $0^{\circ}$ and $90^{\circ}$ represent dislocation segments that are of mixed character. 

\subsubsection{Defining the Transition between Edge and Screw Type}
\label{subsec:transition}
To account for the anisotropy in the Peierls barrier within the PFDD model, Equation \ref{eq:CoreEnergy} will be dependent on the line-character such that dislocations that are of screw type must overcome a much larger energy barrier than dislocation of edge type.  However, the transition of the height of the energy barrier from screw character (maximum) to edge character (minimum) must also be defined in order to allow simulations of general dislocation segments and configurations.  The functional form of such a transition is generally unknown, and the formulation presented here allows for different functions to be tried and tested against atomistic simulations and/or experimental results.  

Perhaps the simplest and easiest definitions of the transition function, $\beta(\theta(\bfx,t))$, are either a simple linear or a sinusoidal transition.  Both of these proposed functional shapes failed to yield matching results to atomistic results (discussed in later sections).  Previous work by Kang \textit{et al.} \cite{Kang2012} included a large number of MD simulations that determined the Peierls stresses of $\bfb=\frac{1}{2}\left[111\right]$ dislocations on the $\left(1\Bar{1}0\right)$ slip plane of BCC Ta as a function of dislocation character.  As expected, pure screw dislocations required the largest stress to initiate motion.  Interestingly, they also found an asymmetric, local maximum in the Peierls stress for a mixed type dislocation with a character angle of $\theta=70.5\degree$. Based on this previous study, we chose to incorporate this second peak (preserving the asymmetry) into the transition function producing a local maximum in the energy barrier for this mixed type dislocation in addition to the global maximum for screw type dislocations.  

The two-peak transition function was initially determined through interpolating the mobility data from Kang \textit{et al.} \cite{Kang2012}. This function is expressed as: 

\begin{equation}
    \label{eq:transiton}
    \beta(\theta) = \begin{cases} 
            1.1603\theta^2 - 2.0431\theta + 1 & \theta\leq 0.39\pi \\
            0.5473\theta^2 - 2.0035\theta + 1.8923 &  0.39\pi<\theta\leq\pi  
            \end{cases}
\end{equation}

\begin{figure}[h]
\centering
\includegraphics[scale=1.0]{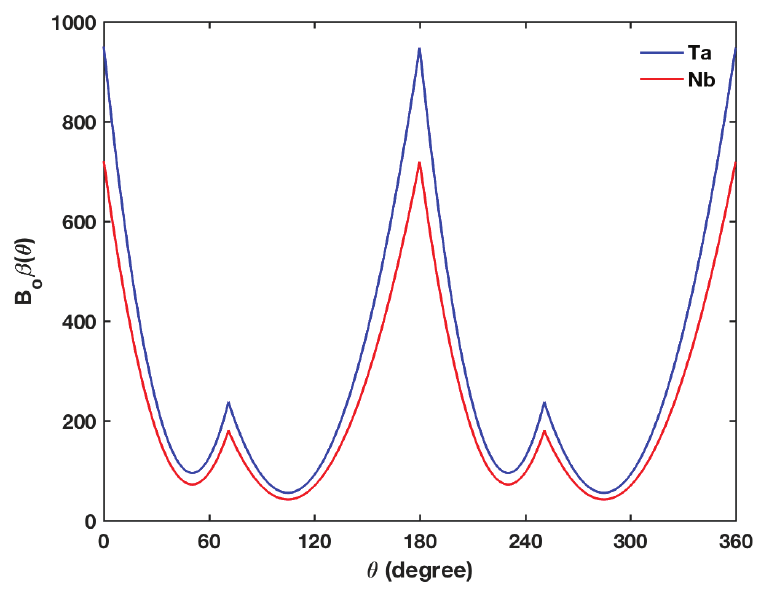}
\caption{\label{fig:BxBeta} Energy barrier $B_o \beta(\theta)$ as a function of line character angle $\theta$ for Ta and Nb.}
\end{figure}

Using this transition function, Equation \ref{eq:CoreEnergy} can be rewritten for BCC materials as:

\begin{equation}
    \label{equ:NewCoreEnergy}
    E^{lattice}(\zeta)=\sum_{\alpha=1}^N\int B_o\beta(\theta^\alpha(\bfx,t))(\sin{n\pi\zeta^\alpha(\bfx,t)})^2\delta_\alpha d^3x,
\end{equation}

\noindent and is now dependent on the character angle, $\theta$.  The parameter $B_o$ is analogous to $B$ originally in Equation \ref{eq:CoreEnergy} in that it parameterizes the magnitude of the energy barrier.  In this case, however, it is different because it will define the magnitude of the energy barrier for only pure screw type dislocations, where the previous parameter $B$ defined the magnitude of the energy barrier for all dislocation character types.  For results presented in later section, $B_o = \gamma_{U}$, where $\gamma_U$ is determined using MS simulations (described next in Section \ref{sec:MSsims}). Unstable stacking fault energies for Ta and Nb are calculated to be 950.21 $mJ/m^2$ and 720.89 $mJ/m^2$, respectively.  The transition of the magnitude of the energy barrier, $B_o\beta(\theta)$, for Ta and Nb are shown in Figure \ref{fig:BxBeta}.

On implementation of this new formulation for $E^{lattice}$ into the TDGL Equation (Equation \ref{eq:TDGL}), we note that an approximation must be made.  The variation of the lattice energy with respect to the order parameter becomes challenging due to the functional dependence of the transition function on the gradient of the order parameter.  Furthermore, $\beta\left(\theta\right)$ is a piecewise function making the variation $\partial E^{lattice}/\partial \zeta$ analytically intractable.  Hence, in calculation of the variation we have made the following approximation,  $\partial E^{lattice}/\partial \zeta = \beta\left(\theta\right)\left(\partial E^{lattice}_{\star}/\partial \zeta\right)$, where $E^{lattice}_{\star}$ is the lattice energy as defined by Equation \ref{eq:CoreEnergy} with $B = B_o$. 

\subsection{Molecular Statics Simulations}
\label{sec:MSsims}

As mentioned in the previous section, the PFDD model was calibrated for Nb and Ta using MS calculations of material parameters such as $\gamma_U$ and the lattice parameter. In addition, isotropic Voigt moduli calculated from the MS predicted stiffness tensor ($\bfC$) were also used to inform the PFDD model. All MS simulations were performed using the LAMMPS\cite{Plimpton1995} software. The components of $\bfC$ were calculated using separate MS simulations in which a 3D periodic 5x5x5 simulation cell, with the Cartesian axes ${x, y, z }$ oriented along the $\left[100\right]$, $\left[010\right]$, and $\left[001\right]$ crystallographic directions, at equilibrium geometry corresponding to P = 0 atm and T = 0 K. This simulation cell was affinely deformed using a prescribed set of lattice strains. These deformed geometries were used to construct energy-strain curves with respect to the applied strain, from which the $C_{ijkl}$ were obtained. The $\gamma_{U}$ were calculated from the $\gamma$-surfaces calculated using the standard procedure,\cite{Vitek:1968, DUESBERY19981481} in which, relaxation of atoms are allowed only in the $\bfm$ direction and periodic boundary conditions are used only in the glide plane. The predicted lattice parameters, linear elastic coefficients, and $\gamma_U$ are provided in Table \ref{tab:table-Matpars}.

\begin{table*}[h!]
  \begin{center}
    \caption{Material parameters calculated from MS simulations and corresponding Young's modulus $E$, shear modulus $G$, and Poisson's ratio $\nu$ calculated with the Voigt isotropic approximation. All elastic coefficients are specified in units of GPa.}
    \label{tab:table-Matpars}
    \begin{tabular}{|c|c|c|c|c|c|c|c|c|c|}
    \hline
     {Material} & {Lattice parameter (\AA)} & C$_{11}$ & C$_{12}$ & C$_{44}$ & $E$ & $G$ & $\nu$ & $\gamma_{U}$ (mJ/m$^2$)\\
      \hline
      Nb & 3.3008 & 246.58 & 133.31 & 28.23 & 110.27 & 39.59 & 0.39 & 720.89 \\
      Ta & 3.3040 & 266.05 & 160.62 & 82.65 & 189.24 & 70.67 & 0.34 & 950.21 \\
      \hline
    \end{tabular}
  \end{center}
\end{table*}

\begin{figure}[h]
\centering
\includegraphics[scale=1.0]{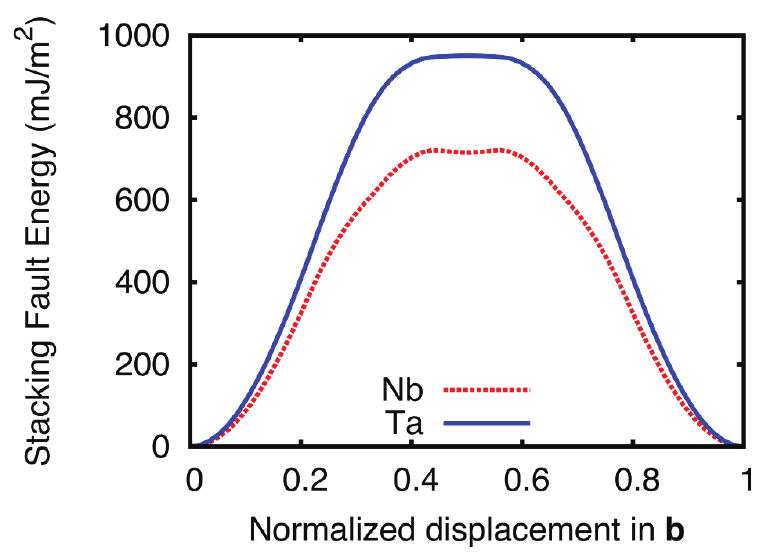}
\caption{\label{fig:SFE} $\left<111\right>$ trace of the $\gamma$-surfaces predicted by MS simulations for Ta and Nb. Note the shallow minima in Nb. }
\end{figure}

The $\left<111\right>$ trace of the $\gamma$-surfaces predicted by MS simulations for Ta and Nb are shown in Figure \ref{fig:SFE}. We used the Finnis-Sinclair inter-atomic potential for Nb and Ta \cite{Ackland2006} to ensure consistency of our simulations with the input parameterization provided to our model by Kang \textit{et al.} \cite{Kang2012}. We note here that a shallow minimum was predicted at ~ $0.5*\left<111\right>$ in the $\gamma$-surface for Nb for the Finnis-Sinclair potential. In addition, cross-slip of screw components during loop expansion was predicted with EAM-type potentials for Ta (Ta1 potential in \cite{PhysRevB.88.134101}) but not with the Finnis-Sinclair potential that was used in the current study.

The PFDD model will be applied to first simulate the expansion of a single dislocation loop (discussed in more detail in the next section).  To validate the PFDD results, MS simulations were carried out for similar cases. Initial simulation cells of size $108\times 39\times 67$ (in units of lattice parameter in the specific crystallographic orientation)  were created using equilibrium lattice parameter corresponding to P = 0 atm and T = 0 K, with the Cartesian axes ${x, y, z }$ oriented along the $\left<111\right>$, $\left<112\right>$, and $\left<110\right>$ crystallographic directions and 3D periodic boundary conditions. Following this a dislocation loop of radius 18*$|\bfb|$ was inserted on a $\left<110\right>$ plane close to the center of the simulation cell using the isotropic linear elasticity displacement field of a dislocation loop, as implemented in the ATOMSK software \cite{HIREL2015212}. A RSS was applied to expand the loop on the slip plane in the $\bfb$ direction. The RSS was applied by applying an affine deformation to all the atoms in the simulation cell to result in the strain corresponding to the RSS predicted by a linear elasticity model. Extraction and visualization of the dislocation loops were done using OVITO \cite{Stukowski2009, Stukowski2012, Stukowski2014}.

\section{Results and Discussion}
\label{sec:results}

To test the extension of the PFDD model described above, we consider two different simulation configurations. The first is expansion of a single, initially circular, dislocation loop and the second is propagation of a kink-pair.  The first case is informative because in BCC metals the dislocation loop will not expand symmetrically due to the difference in the energy barrier for the screw, edge, and mixed typed dislocations. A dislocation loop contains the full range of line characters, hence the role played by the character-dependent core energy on dislocation motion can be assessed. 

\begin{figure*}[h]
    \centering
    \includegraphics[scale=1.0]{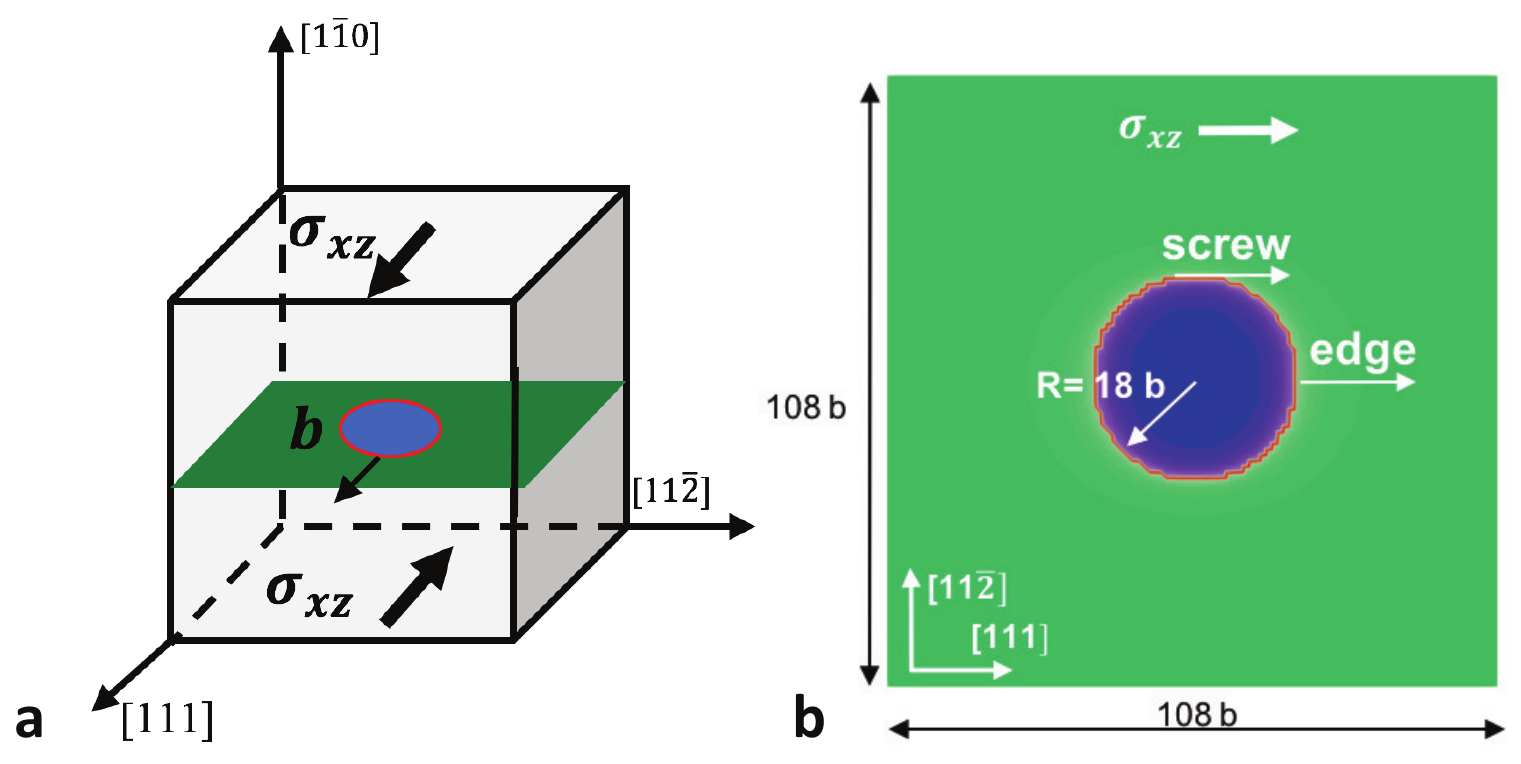}
    \caption{Figure (a) presents a schematic of the 3D simulation cell for the PFDD simulations of expansion of a single dislocation loop.  Figure (b) presents the initial conditions on the glide plane for these simulations.}%
    \label{fig:setupLoop}%
\end{figure*}

An example of the initial conditions for the expansion of a loop in PFDD is shown in Figure \ref{fig:setupLoop}.  The 3D configuration is shown schematically in Figure \ref{fig:setupLoop}(a).  This consists of a cubic box with dimensions  $108\times108\times108$ (in units of $|\bfb|$) and an applied stress, $\sigma_{xz}$, that will cause the loop to expand. Figure \ref{fig:setupLoop}(b) shows a close-up of the loop on the glide plane as generated from PFDD. A circular dislocation loop with radius 18*$|\bfb|$ is initialized on the $(1\Bar{1}0)$ slip plane with $\bfb=\frac{1}{2}[111]$.  This loop size was chosen to match the loop radius used in the MS simulations.  The green region in Figure \ref{fig:setupLoop}(b) indicates areas that have not been slipped, whereas the blue region are areas that have been slipped by a single perfect dislocation.  The red line represents the dislocation line. 

\begin{figure*}[h]
    \centering
    \includegraphics[scale=1.0]{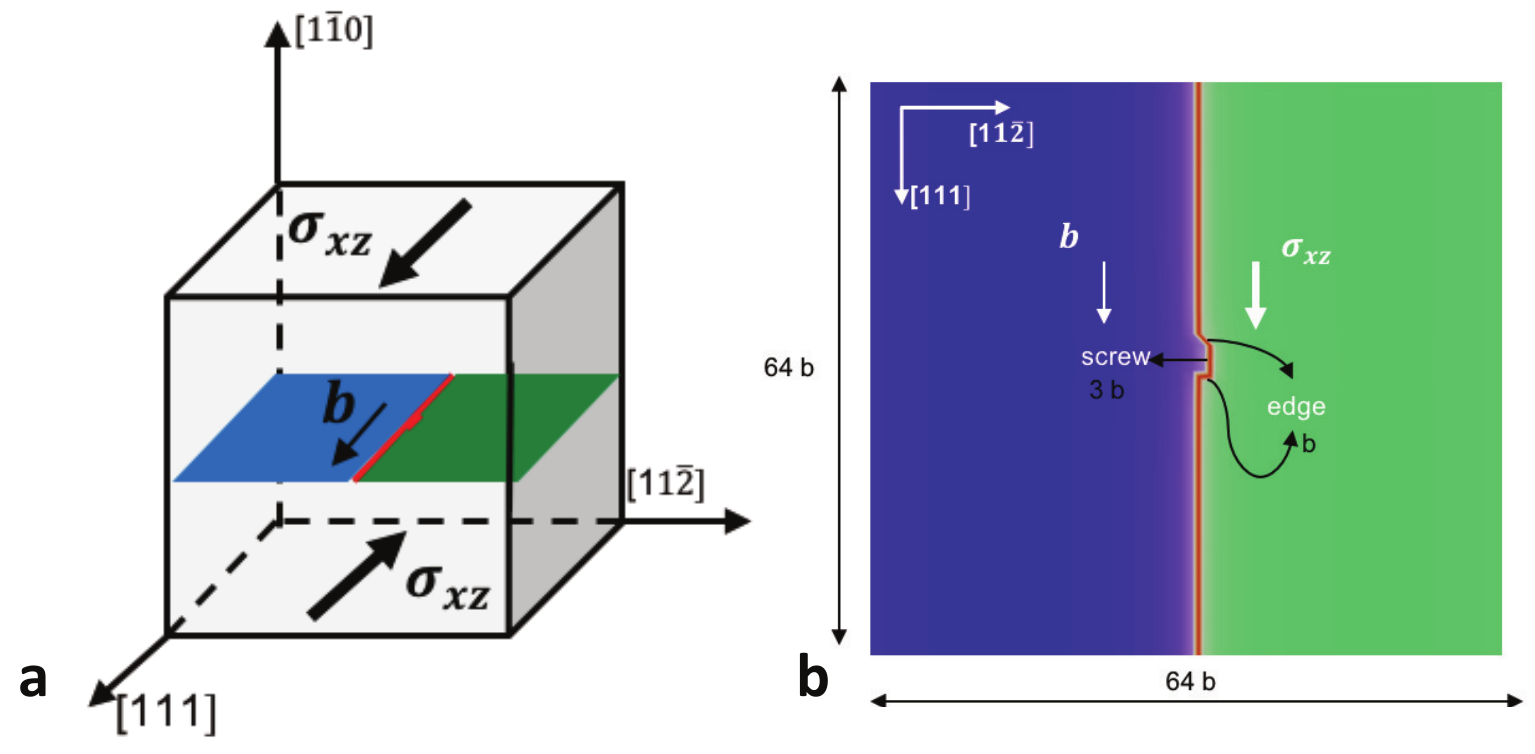}
    \caption{Figure (a) presents a schematic of the 3D simulation cell for the PFDD simulations propagation of a kink pair.  Figure (b) presents the initial conditions on the glide plane for this simulation.}%
    \label{fig:setupKink}%
\end{figure*}

The second simulation configuration is for kink-pair propagation.  Kink-pairs are a common mechanism for screw dislocation motion in which the screw dislocation nucleates a small step in the dislocation line.  This step is comprised of two oppositely signed edge segments and one screw segment.  Due to the ease at which edge dislocations move, the two edge segments will move away from each other (with appropriate loading conditions) along the length of the straight screw dislocation.  This, in turn, propagates the screw dislocation.  Figure \ref{fig:setupKink} presents the PFDD simulation set-up for the kink-pair simulation. In this work, we do not explicitly address kink-pair nucleation.  Hence, the kink-pair is placed along the straight screw dislocation as part of the initial conditions, and allowed to propagate under an applied load.

All simulations had a computational grid spacing of 1*$|\bfb|$ in the $x$, $y$, and $z$ directions.  In addition, due to the use of a Fourier transform in the calculation of Equation \ref{eq:Eelast}, all PFDD simulations have periodic boundary conditions.  This will result in the presence of some image forces due to dislocations in neighboring periodic cells, which is also the case in the MS simulations.  In both methods, the interactions with image dislocations will have an affect on the simulation, particularly when dislocation line are near the edges of the simulation cell.  Finally, in these PFDD simulations the materials are assumed to be elastically isotropic.  The Voigt isotropic moduli are reported in Table \ref{tab:table-Matpars}.

\subsection{Expansion of a Perfect Dislocation Loop}

To test our character-dependent lattice energy, we first simulated the expansion of a perfect dislocation loop in both PFDD and MS for Ta and also Nb. Ta provides the most direct comparison between modeling approaches since the functional form of the transition function is formulated using MD information generated for Ta.  In addition, these MD results utilized the same interatomic potential as the MS results presented here. However, the model should be general enough to extend to other BCC metals.  Hence, we have also modeled Nb using the same piecewise function for the transition function.  This is somewhat of an extrapolation, since Peierls stress or Peierls energy barrier calculations as a function character have not been previously done for Nb.

\begin{figure*}[h!]
\centering
\includegraphics[scale=1.0]{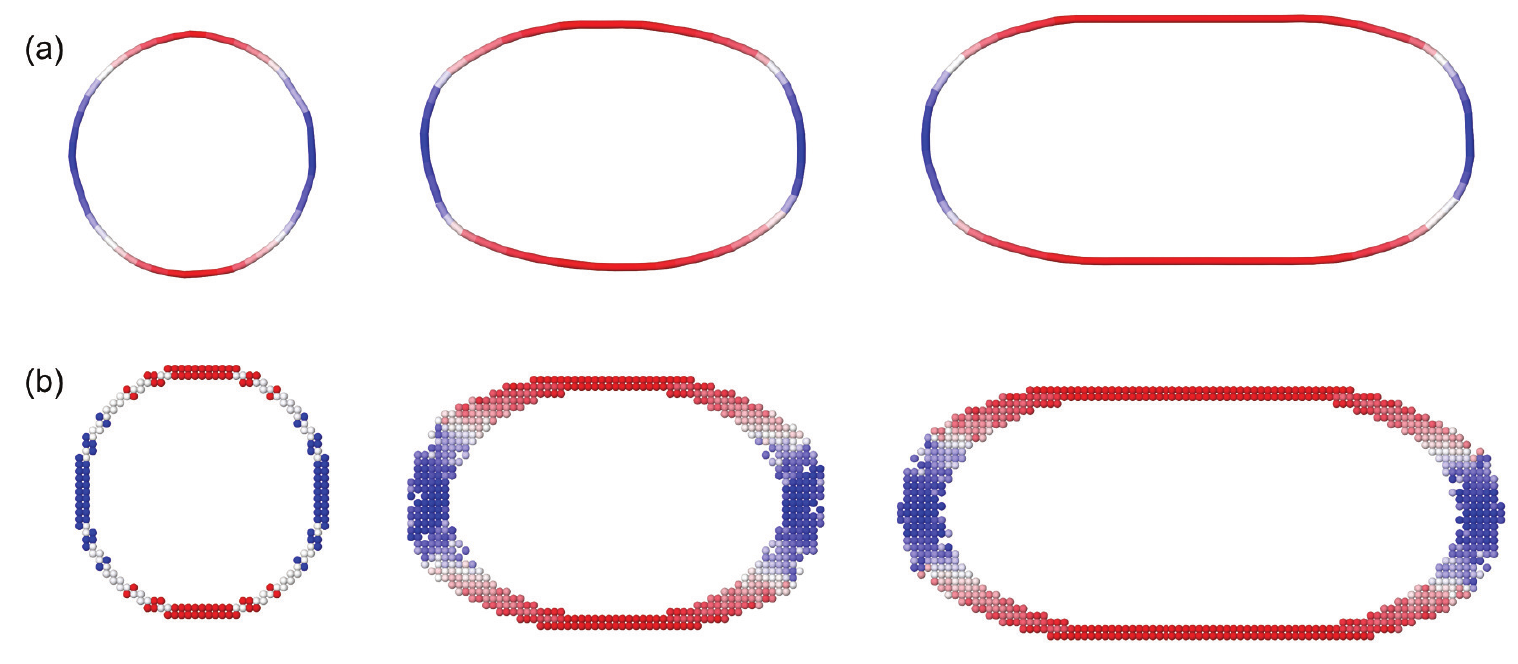}
\caption{\label{fig:TaLoopExp} Ta loop expansion snapshots in (a) MS (b) PFDD at normalized time $t=0, 0.2, 0.4$. The time for each snapshots is normalized with respect to the time that loops reach their steady state.  The loops are colored by character with screw-type dislocation shown in red, and edge-type dislocations shown in blue.}
\end{figure*}

\begin{figure*}[h!]
\centering
\includegraphics[scale=1.0]{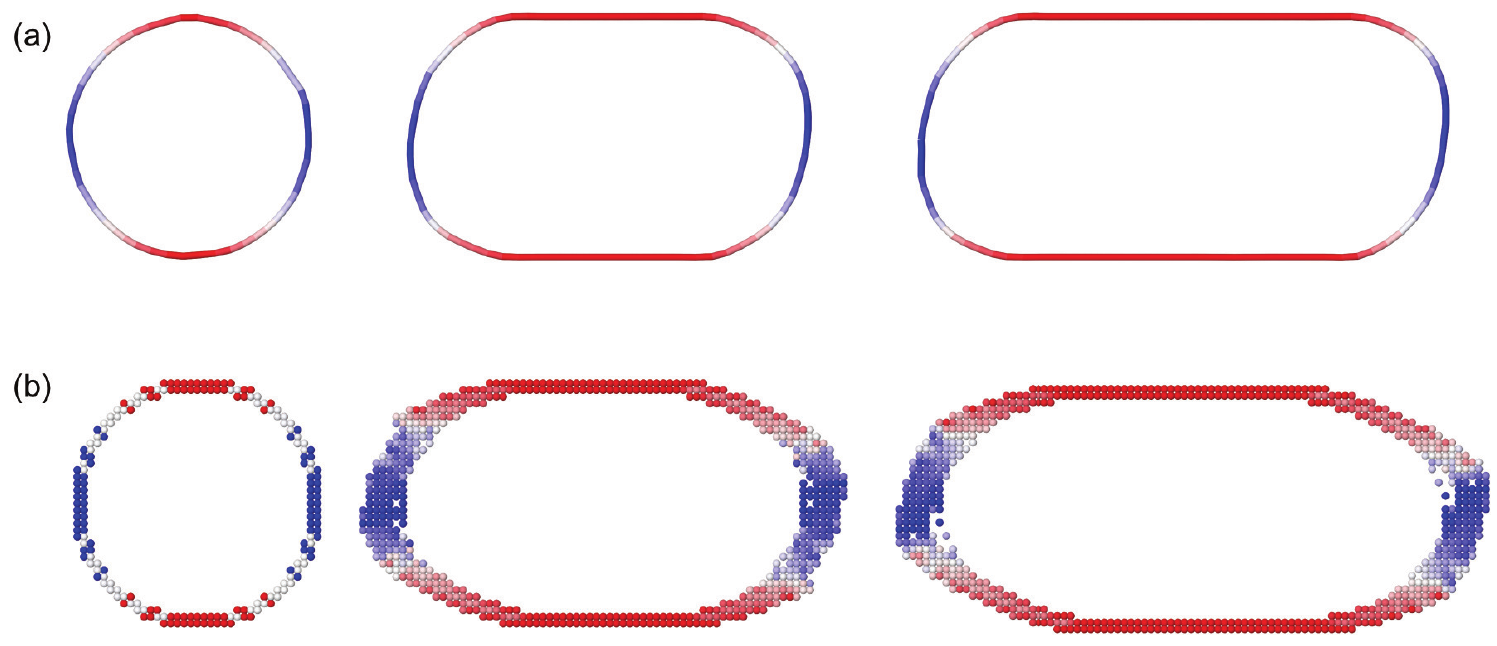}
\caption{\label{fig:NbLoopExp} Nb loop expansion snapshots in (a) MS (b) PFDD at normalized time $t=0, 0.2, 0.4$. Time for each snapshots is normalized with respect to the time that loops reach their steady state.  The loops are colored by character with screw-type dislocation shown in red, and edge-type dislocations shown in blue.}
\end{figure*}

Figures \ref{fig:TaLoopExp} and \ref{fig:NbLoopExp} show the expanding dislocation loop at different stages of the minimization for an applied RSS. A RSS value of 2 GPa was used for the Ta and Nb in both the PFDD and MS simulations. This value was selected to be smaller than the Peierls stress for the screw dislocation, but large enough to overcome the elastic attraction between the dislocation segments within the loop. The loops are colored by character, with a continuous red-white-blue coloring scheme transitioning from screw (red) to edge (blue) character. For all quantitative comparisons, screw dislocations were defined as $\theta = 0\degree\pm 10\degree$ and edge as $\theta = 90\degree\pm 10\degree$.

Figures \ref{fig:TaLoopExp} and \ref{fig:NbLoopExp} show that, with the addition of line character dependence to the $E^{lattice}$ term, PFDD qualitatively captures the correct loop growth as predicted by MS. Rather than symmetric expansion of the loop, the edge segments propagate while the screw dislocation segments remain relatively stationary and increase in length. This quickly results in an elliptical loop shape.  The edge segments continue to propagate until they annihilate with the periodic image dislocations, leaving two long pure screw dislocation segments.  This final configuration is considered to be the steady state solution for these simulations. 

\begin{figure*}[h!]
    \centering
    \includegraphics[scale=1.0]{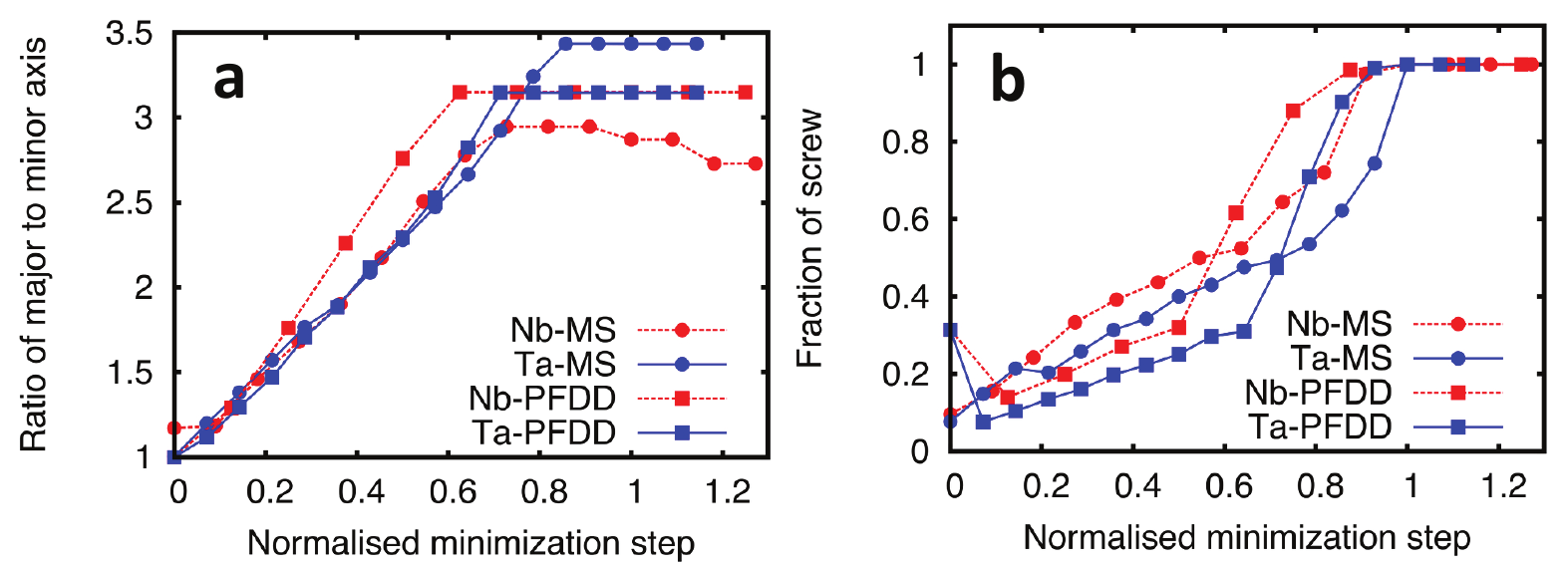}
    \caption{Quantitative comparisons between PFDD and MS for Ta and Nb: (a) Ratio of major to minor axis of dislocation loop (b) Fraction of screw segment.}%
    \label{fig:QuanComp}%
\end{figure*}

Note that, in Figure \ref{fig:NbLoopExp}, MS predicts a skewed or slanted loop structure for Nb due to the asymmetric peak in the Peierls barrier at mixed type dislocations with $\theta=70.5\degree$, which does not exist for segments with $\theta \approx 135\degree$. Hence, while the $70.5\degree$ segments move more slowly, the $\sim 135\degree$ segments are not restricted in such a way and can move more quickly that their $70.5\degree$ counterparts producing a loop structure that appears skewed or slanted. This is also qualitatively captured by PFDD as the line-character dependent $E^{lattice}$ includes multiple minima as discussed in section \ref{subsec:transition}.  This skewed loop structure is slightly more pronounced in the MS results that the PFDD.  This may be due in part to the assumption of elastic isotropy used in the PFDD simulations.  In addition, due to the lower Peierls stress and weaker elastic interactions, the segments at  $\sim 135\degree$ move more in Nb compared to Ta, at the same RSS of 2 GPa. This makes the skewed loop structure prominent in Nb compared to Ta. A more quantitative comparison between the MS and PFDD simulations of dislocation loop expansion is shown in Figure \ref{fig:QuanComp}. The comparisons are made at different stages of minimization, for which the minimization steps were normalized based on the step at which edge components of the loop interact across the periodic boundaries and annihilate each other.

Figure \ref{fig:QuanComp}(a) compares the ratio between major and minor axis of the loop. Effectively this metric measures the ratio of the distance between opposite screw components to the distance between the opposite edge components. This ratio is initially close to 1, as the loop is approximately circular, and quickly changes to a value > 1 as the loop expands asymmetrically. A fair agreement is observed between PFDD and MS predictions, with the ratio saturating to a value $\approx$ 3. 

The PFDD values more closely match the MS results for Ta.  Any differences seen in this case lie primarily in the steady state result, where PFDD slightly under-predicts the MS result.  This discrepancy is likely due to the differences in the elastic interaction between the screw type dislocations as captured in MS versus PFDD.  The model assumes the material is isotropic and, therefore, the screw and edge segments within the same loop do not interact elastically.  As a measure of elastic anisotropy, Zener anisotropy ratios, $A = \frac{2C_{44}}{C_{11}-C_{12}}$, can be calculated as: $A_{Ta} = 1.568$ and $A_{Nb} = 0.499$.  Although the departure from unity (signifying isotropy) in the anisotropy ratios is similar for Ta and Nb,$A_{Nb} < 1$ and $A_{Ta}>1$.  In the case of Nb, PFDD shows a very good match in the aspect ratio at the initial stages of loop growth, but quickly starts to over-predict the MS results.  Converse to Ta, PFDD over-predicts the steady result calculated by MS for Nb.  

Figure \ref{fig:QuanComp}(b) compares the fraction of the screw dislocation in the expanding loop at various stages of minimization. The fraction was determined by counting all dislocation segments with a line-character $\theta = 0\degree\pm 10\degree$ as screw and comparing this to  total number of dislocation segments at various stages of minimization. Note that beyond the initial few steps in minimization, both MS and PFDD predict that the expanding loop has a higher screw fraction in Nb compared to Ta.  About halfway through the simulation, there is a noticeable change in slope indicating a sudden growth in screw type dislocations, particularly in the PFDD generated results.  This change in slope indicates where elastic interactions with image dislocations start to dominate, motivating the edge type dislocations to annihilate with the neighboring dislocations.  This change in slope is present in the MS simulations, although it is more gradual and occurs later in the normalized simulation time.

The PFDD and MS results show reasonable agreement.  Interestingly the PFDD results first under-predict the fraction of screw dislocation in the loop, and then over-predict the amount of screw dislocation once the image forces start becoming more dominant.  We note that in the initial configuration of the loop in PFDD, there are relatively long screw segments due to the cubic computational grid.  Hence, initially the PFDD results show a decrease in the fraction of screw dislocation as the loop quickly relaxes to minimize the amount of screw type dislocation once energy minimization occurs.  

\subsection{Propagation of a Straight Screw Dislocation through Kink-Pair Motion}

\begin{figure*}[h!]
\centering
\includegraphics[scale=1.0]{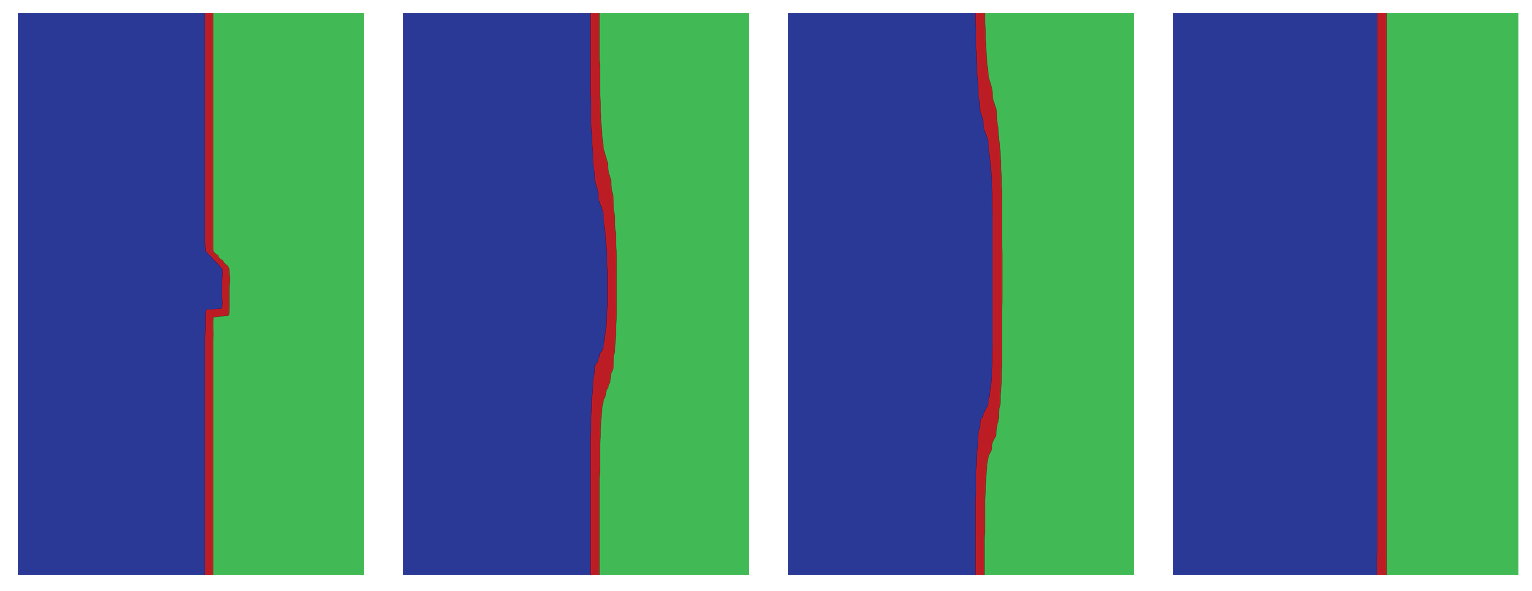}
\caption{\label{fig:TaKink} PFDD simulation of screw dislocation motion by propagation of kinks with edge character. Snapshots are shown at normalized time, $t=0, 0.2, 0.3, 1.0$.}
\end{figure*}

In addition to the expansion of a loop, the propagation of a straight screw dislocation through kink-pair motion in Ta was also modeled in PFDD.  Simulation results are shown in Figure \ref{fig:TaKink}.  The kink-pair is initially placed along the screw dislocation.  With an applied stress of 1.77 GPa, the edge segments of the kink-pair rapidly propagate to the end of the simulation cell where they annihilate with the image dislocation in the periodic cells.  The screw type-dislocation has then propagated forward by on Burgers vector step.

\section{Conclusions}
\label{sec:conclusions}

We have presented a PFDD model extended to BCC metals.  In particular, the model extensions are focused on capturing the dependence of the Peierls energy barrier on character.  This has been achieved through the addition of a transition function, which describes how the energy for a dislocation to glide through the crystal lattice depends on the dislocation character angle, $\theta$.  In this work, the functional form of the energy barrier transition has been informed by MD calculations of Peierls stress as a function character for Ta previously reported in \cite{Kang2012}.  We realize that this approach relies on the quality of the interatomic potential used for the MD simulations, and we note that the form of the transition function could easily be modified as new atomisitic data becomes available.  

The character dependent PFDD model was demonstrated by modeling expansion of a dislocation loop and propagation of a screw dislocation through kink-pair motion in Ta.  In addition, a dislocation loop expansion was modeled in Nb to illustrate the ability of the transition function to extrapolate to other BCC metals beyond Ta.  The PFDD results for expansion of a dislocation loop in both Ta and Nb were directly compared to similar simulations completed with MS.  The results from the two methods compared reasonably well, with the largest source or error likely being in the calculation of the elastic interactions, which were considered only isotropically in PFDD.  Perhaps most notable were the evolution of a skewed loop structure and the higher screw fraction in Nb as shown in the MS calculations, which are also captured by PFDD.   

Ultimately, dislocation dynamics in BCC metals are rate dependent, giving rise to a macroscopic deformation response that is highly sensitive to the applied strain rate and temperature.  With the mechanistic model in place in the present framework, it is now possible to incorporate temperature and rate effects on dislocation motion. With such additions, this model could model both nucleation and propagation of kink-pairs for example.  Such topics are interesting extensions for future work.

\section*{Data Availability}

Data is available from the authors upon request.
	
\section*{Acknowledgments}

XP and KD acknowledge support from ONR Applied and Computational Analysis (N00014-14-1-0715 and N00014-18-1-2528), NSF DMREF program (1628994), and NSF ACI program (1548562, 1445606, and TG-DMR120046). NM gratefully acknowledges support from the U.S. Department of Energy through the LANL/LDRD Program and the Center for Nonlinear Studies (CNLS) for this work. IJB gratefully acknowledges support in part from the Office of Naval Research under contract ONR BRC Grant N00014-18-1-2392. AH gratefully acknowledges support from the Materials project within the Physics and Engineering Models (PEM) Subprogram element of the Advanced Simulation and Computing (ASC) Program at Los Alamos National Laboratory (LANL). This manuscript has been assigned LA-UR-19-24421.


\bibliographystyle{alpha}
\bibliography{references}

\newcommand{\etalchar}[1]{$^{#1}$}
\begin{thebibliography}{WFVdGN02}

\bibitem[ACT{\etalchar{+}}07]{Arsenlis:2007}
A.~Arsenlis, W.~Cai, M.~Tang, M.~Rhee, T.~Oppelstrup, G.~Hommes, T.~G. Pierce,
  and V.~V. Bultav.
\newblock Enabling strain hardening simulations with dislocation dynamics.
\newblock {\em Modelling and Simulation in Materials Science and Engineering},
  15:553--595, 2007.

\bibitem[ALM01]{Argaman:2001}
N.~Argaman, O.~Levy, and G.~Makov.
\newblock When do 2-d dislocations form cellular structures?
\newblock {\em Materials Science and Engineering A}, 309-310:386--392, 2001.

\bibitem[AT06]{Ackland2006}
G~J Ackland and R~Thetford.
\newblock {An improved N -body semi-empirical model for body-centred cubic
  transition metals}.
\newblock {\em Philosophical Magazine A}, 8610, 2006.

\bibitem[BC06]{BulatovCai:2006}
V.~V. Bulatov and W.~Cai.
\newblock {\em Computer Simulations of Dislocations}.
\newblock Oxford University Press, Great Britain, 2006.

\bibitem[BH67]{Byron:1967}
J.~F. Byron and D.~Hull.
\newblock Plastic deformation of tantalum single crystals. i. the surface
  morphology of yield.
\newblock {\em Journal of Less-Common Metals}, 13:71--84, 1967.

\bibitem[BH16]{Beyerlein2016}
I~J Beyerlein and A~Hunter.
\newblock {Understanding nanoscale dislocation mechanics using phase field
  dislocation dynamics}.
\newblock {\em Philos. Trans. R. Soc. A}, 374(2066):20150166, 2016.

\bibitem[BT67]{Bowen:1967}
J.~W. Bowen, D. K.~Christian and G.~Taylor.
\newblock Deformation properties of niobium single crystals.
\newblock {\em Canadian Journal of Physics}, 45(2):903--938, 1967.

\bibitem[BTB{\etalchar{+}}14]{Bertin:2014}
N.~Bertin, C.~N. Tom\'{e}, I.~J. Beyerlein, M.~R. Barnett, and L.~Capolungo.
\newblock On the strength of dislocation interactions and their effect on
  latent hardening in pure magnesium.
\newblock {\em International Journal of Plasticity}, 62:72--92, 2014.

\bibitem[CB04]{Cai2:2004}
W.~Cai and V.~V. Bulatov.
\newblock Mobility laws in dislocation dynamics simulations.
\newblock {\em Materials Science and Engineering A}, 387-389:277--281, 2004.

\bibitem[CHBK15]{Cao:2015}
L.~Cao, A.~Hunter, I.~J. Beyerlein, and M.~Koslowski.
\newblock The role of partial mediated slip during quasi-static deformation of
  3d nanocrystalline metals.
\newblock {\em Journal of the Mechanics and Physics of Solids}, 78:415--426,
  2015.

\bibitem[Chr83]{Christian:1983}
J.~W. Christian.
\newblock Some surprising feature of the plastic deformation of body-centered
  cubic metals and alloys.
\newblock {\em Metallurgical Transactions A}, 14A:1237--1256, 1983.

\bibitem[CSPK17]{Cao:2017}
L.~Cao, A.~Sengupta, D.~Pantuso, and M.~Koslowski.
\newblock Effect of texture and grain size on the residual stress of
  nanocrystalline thin films.
\newblock {\em Modelling and Simulation in Materials Science and Engineering},
  25:075004, 2017.

\bibitem[DK97]{Devincre:1997}
B.~Devincre and L.~P. Kubin.
\newblock Mesoscopic simulations of dislocations and plasticity.
\newblock {\em Materials Science and Engineering A}, 8:14, 1997.

\bibitem[DRC{\etalchar{+}}16]{Dezerald:2016}
L.~Dezerald, D.~Rodney, E.~Clouet, L.~Ventelon, and F.~Willaime.
\newblock Plastic anisotropy and dislocation trajectory in bcc metals.
\newblock {\em Nature Communications}, 7:11695, 2016.

\bibitem[DSDR10]{Dmitrieva:2010}
O.~Dmitrieva, J.~V. Svirina, E.~Demir, and D.~Raabe.
\newblock Investigation of the internal substructure of microbands in a
  deformed copper single crystal: experiments and dislocation dynamics
  simulation.
\newblock {\em Modelling and Simulation in Materials Science and Engineering},
  18:085011, 2010.

\bibitem[DV98a]{Duesbery:1998}
M.~S. Duesbery and V.~Vitek.
\newblock Plastic anisotropy in bcc transition metals.
\newblock {\em Acta Materialia}, 46(5):1481--1492, 1998.

\bibitem[DV98b]{DUESBERY19981481}
M.S. Duesbery and V.~Vitek.
\newblock Plastic anisotropy in b.c.c. transition metals.
\newblock {\em Acta Materialia}, 46(5):1481 -- 1492, 1998.

\bibitem[EABG08]{ElAwady:2008}
J.~A. El-Awady, S.~B. Biner, and N.~M. Ghoniem.
\newblock A self-consistent boundary element, parametric dislocation dynamics
  formulation of plastic flow in finite volumes.
\newblock {\em Journal of the Mechanics and Physics of Solids},
  56(5):2019--2035, 2008.

\bibitem[EAFH16]{ElAwady:2016}
J.~A. El-Awady, H.~Fan, and A.~M. Hussein.
\newblock Advances in discrete dislocation dynamics modeling of size-affected
  plasticity.
\newblock In C.~Weinberger and G.~Tucker, editors, {\em Multiscale Materials
  Modeling for Nanomechanics, Springer Series in Materials Science}, volume
  245. Springer, 2016.

\bibitem[GFMH15]{Gao:2015}
S.~Gao, M.~Fivel, A.~Ma, and A.~Hartmaier.
\newblock {Influence of misfit stresses on dislocation glide in single crystal
  superalloys: a three-dimensional discrete dislocation dynamics study}.
\newblock {\em Journal of the Mechanics and Physics of Solids}, 76:276, 2015.

\bibitem[GTS00]{Ghoniem:2000}
N.~M. Ghoniem, S.~H. Tong, and L.~Z. Sun.
\newblock Parametric dislocation dynamics: A thermodynamics-based approach to
  investigations of mesoscopic plastic deformation.
\newblock {\em Physical Review B}, 61(2):913--927, 2000.

\bibitem[GV19]{Groger:2019}
R.~Gr{\"{o}}ger and V.~Vitek.
\newblock Impact of non-schmid stress components present in the yield criterion
  for bcc metals on the activity of \{110\}<111> slip systems.
\newblock {\em Computational Materials Science}, 159:297--305, 2019.

\bibitem[HB01]{Hull_Bacon:2001}
D.~Hull and D.J. Bacon.
\newblock {\em Introduction to dislocations}.
\newblock Elsevier Butterworth-Heinemann, Oxford, 2001.

\bibitem[HB14a]{Hunter3:2014}
A.~Hunter and I.~J. Beyerlein.
\newblock Predictions of an alternative pathway for grain-boundary driven
  twinning.
\newblock {\em Applied Physics Letters}, 104(233112):1--4, 2014.

\bibitem[HB14b]{Hunter:2014}
A.~Hunter and I.~J. Beyerlein.
\newblock Stacking fault emission from grain boundaries: Material dependencies
  and grain size effects.
\newblock {\em Materials Science and Engineering A}, 600:200--210, 2014.

\bibitem[HB15]{Hunter:2015}
A.~Hunter and I.~J. Beyerlein.
\newblock Relationship between monolayer stacking faults and twins in
  nanocrystals.
\newblock {\em Acta Materialia}, 88:207--217, 2015.

\bibitem[HBGK11]{Hunter:2011}
A.~Hunter, I.~J. Beyerlein, T.~C. Germann, and M.~Koslowski.
\newblock Infuence of the stacking fault energy surface on partial dislocations
  in fcc metals with a three-dimensional phase field model.
\newblock {\em Physical Review B}, 84(144108):1--10, 2011.

\bibitem[Hir15]{HIREL2015212}
Pierre Hirel.
\newblock Atomsk: A tool for manipulating and converting atomic data files.
\newblock {\em Computer Physics Communications}, 197:212 -- 219, 2015.

\bibitem[HL68]{Hirth:1968}
J.~P. Hirth and J.~Lothe.
\newblock {\em Theory of Dislocations}.
\newblock McGraw-Hill, New York, 1968.

\bibitem[HLB18]{Hunter:2018}
A.~Hunter, B.~Leu, and I.~J. Beyerlein.
\newblock A review of slip transfer: applications of mesoscale techniques.
\newblock {\em J. Mater. Sci.}, 53:5584--5603, 2018.

\bibitem[HRU{\etalchar{+}}17]{Hussein:2017}
A.~M. Hussein, S.~I. Rao, M.~D. Uchic, T.~A. Parthasarathy, and J.~A. El-Awady.
\newblock {The strength and dislocation microstructure evolution in superalloy
  microcrystals}.
\newblock {\em Journal of the Mechanics and Physics of Solids}, 99:146, 2017.

\bibitem[Hsi10]{Hsiung:2010}
L.~L. Hsiung.
\newblock On the mechanism of anomalous slip in bcc metals.
\newblock {\em Materials Science and Engineering A}, 528:329--337, 2010.

\bibitem[HZB14]{Hunter2:2014}
A.~Hunter, R.~F. Zhang, and I.~J. Beyerlein.
\newblock The core structure of dislocation and their relationship to the
  material $\gamma$-surface.
\newblock {\em Journal of Applied Physics}, 115:134314, 2014.

\bibitem[HZT12]{Huang:2012}
M.~Huang, L.~Zhao, and J.~Tong.
\newblock Discrete dislocation dynamics modelling of mechanical deformation of
  nickel-based single crystal superalloys.
\newblock {\em International Journal of Plasticity}, 28:141--158, 2012.

\bibitem[JD97]{Joos:1997}
B.~Jo$\acute{o}$s and M.~S. Duesbery.
\newblock The peierls stress of dislocations: An analytic formula.
\newblock {\em Physical Review Letters}, 78(2):266--269, 1997.

\bibitem[KBC12]{Kang2012}
K.~Kang, V.~V. Bulatov, and W.~Cai.
\newblock {Singular orientations and faceted motion of dislocations in
  body-centered cubic crystals}.
\newblock {\em Proceedings of the National Academy of Sciences},
  109(38):15174--15178, 2012.

\bibitem[KC92]{Kubin2:1992}
L.~P. Kubin and G.~Canova.
\newblock The modelling of dislocation patterns.
\newblock {\em Scripta Metallurgica et Materialia}, 27:957--962, 1992.

\bibitem[KCC{\etalchar{+}}92]{Kubin:1992}
L.~P. Kubin, G.~Canova, M.~Condat, B.~Devincre, V.~Pontikis, and Yves
  Br{\'{e}}chet.
\newblock Dislocation microstructures and plastic flow: A 3d simulation.
\newblock In {\em Non Linear Phenomena in Materials Science II}, volume~23 of
  {\em Solid State Phenomena}, pages 455--472. Trans Tech Publications, 1 1992.

\bibitem[KCO02]{Koslowski2002}
M.~Koslowski, A.~M. Cuiti{\~{n}}o, and M.~Ortiz.
\newblock {A phase-field theory of dislocation dynamics, strain hardening and
  hysteresis in ductile single crystals}.
\newblock {\em Journal of the Mechanics and Physics of Solids},
  50(12):2597--2635, 2002.

\bibitem[KDT98]{Kubin:1998}
L.~P. Kubin, B.~Devincre, and M.~Tang.
\newblock Mesoscopic modelling and simulation of plasticity in fcc and bcc
  crystals: Dislocation intersection and mobility.
\newblock {\em Journal of Computer-Aided Materials Design}, 5:31--54, 1998.

\bibitem[KM03]{Kocks:2003}
U.~F. Kocks and H.~Mecking.
\newblock Physics and phenomenology of strain hardening: the fcc case.
\newblock {\em Progress in Material Sciences}, 48(3):171--273, 2003.

\bibitem[Kub93]{Kubin:1993}
L.~P. Kubin.
\newblock Dislocation patterning during multiple slip of fcc crystals: A
  simulation approach.
\newblock {\em Physica Status Solidi A}, 135:433--443, 1993.

\bibitem[KWLL11]{Koslowski:2011}
M.~Koslowski, D.~Wook~Lee, and L.~Lei.
\newblock Role of grain boundary energetics on the maximum strength of nano
  crystalline nickel.
\newblock {\em Journal of the Mechanics and Physics of Solids}, 59:1427--1436,
  2011.

\bibitem[LBW18]{Lim:2018}
H.~Lim, C.~C. Battaile, and C.~R. Weinberger.
\newblock Simulating dislocation plasticity in bcc metals by integrating
  fundamental concepts with macroscale models.
\newblock In M.~F. Horstemeyer, editor, {\em Integrated Computational Materials
  Engineering (ICME) for Metals: Concepts and Case Studies}, chapter~4, pages
  71--106. John Wiley \& Sons, 2018.

\bibitem[LCM{\etalchar{+}}19]{Lim:2019}
H.~Lim, J.~D. Carroll, J.~R. Michael, C.~C. Battaile, S.~R. Chen, and J.~M.~D.
  Lane.
\newblock Investigating active slip planes in tantalum.
\newblock {\em Acta Materialia}, 2019.

\bibitem[LeS13]{Lesar:2013}
R.~LeSar.
\newblock {\em Introduction to Computational Materials Science Fundamentals to
  Applications}.
\newblock Cambridge University Press, United Kingdom, 2013.

\bibitem[LK79]{Louchet:1979}
F.~Louchet and L.~P. Kubin.
\newblock In situ deformation of b.c.c. crystals at low temperatures in a
  high-voltage electron microscope: Dislocation mechanisms and strain-rate
  equation.
\newblock {\em Philosophical Magazine A}, 39:433--454, 1979.

\bibitem[LMK13]{Lei:2013}
L.~Lei, J.~L. Marian, and M.~Koslowski.
\newblock Phase-field modeling of defect nucleation and propagation in domains
  with material inhomogeneities.
\newblock {\em Modelling and Simulation in Materials Science and Engineering},
  21(025009):1--15, 2013.

\bibitem[LT62]{Low:1962}
J.~R. Low and A.~M. Turkalo.
\newblock Slip band structure and dislocation multiplication in silicon-iron
  crystals.
\newblock {\em Acta Metallurgica}, 10:215--227, 1962.

\bibitem[LTBL17]{Louchez:2017}
M.~A. Louchez, L.~Thuinet, R.~Besson, and A.~Legris.
\newblock Microscopic phase-field modeling of hcp|fcc interfaces.
\newblock {\em Computational Materials Science}, 132:62--73, 2017.

\bibitem[MDK02]{Madec2:2002}
R.~Madec, B.~Devincre, and L.~P. Kubin.
\newblock Simulation of dislocation patterns in multislip.
\newblock {\em Scripta Materialia}, 47:689--695, 2002.

\bibitem[MDK04]{Monnet:2004}
G.~Monnet, B.~Devincre, and L.~P. Kubin.
\newblock Dislocation study of prismatic slip systems and their interactions in
  hexagonal close packed metals: application to zirconium.
\newblock {\em Acta Materialia}, 52:4317--4328, 2004.

\bibitem[MKO11]{Mori:2011}
H.~Mori, H.~Kimizuka, and S.~Ogata.
\newblock Microscropic phase-field modeling of edge and screw dislocation core
  structures and peierls stresses of bcc iron.
\newblock {\em J. Japan Inst. Metals}, 75(2):104--109, 2011.

\bibitem[Mur87]{mura1987general}
Toshio Mura.
\newblock General theory of eigenstrains.
\newblock In {\em Micromechanics of defects in solids}, pages 1--73. Springer,
  1987.

\bibitem[Pei40]{Peierls:1940}
R.~Peierls.
\newblock The size of a dislocation.
\newblock {\em Proceedings of the Physical Society}, 52(1):24--37, 1940.

\bibitem[Pliov]{Plimpton1995}
S.~Plimpton.
\newblock {Fast parallel algorithms for short-range molecular dynamics}.
\newblock {\em Journal of Computational Physics}, 117:1--19, 1995; Also see
  http://lammps.sandia.gov/.

\bibitem[QZS{\etalchar{+}}19]{Qiu:2019}
D.~Qiu, P.~Zhao, C.~Shen, W.~Lu, D.~Zhang, M.~Mrovec, and Y.~Wang.
\newblock Predicting grain boundary structure and energy in bcc metals by
  integrated atomistic and phase-field modeling.
\newblock {\em Acta Materialia}, 164:799--809, 2019.

\bibitem[RGG{\etalchar{+}}13]{PhysRevB.88.134101}
R.~Ravelo, T.~C. Germann, O.~Guerrero, Q.~An, and B.~L. Holian.
\newblock Shock-induced plasticity in tantalum single crystals: Interatomic
  potentials and large-scale molecular-dynamics simulations.
\newblock {\em Physical Review B}, 88:134101, Oct 2013.

\bibitem[Ric71]{Richter:1971}
J.~Richter.
\newblock The influence of temperature on slip behavior of molybdenum single
  crystals deformed in tension in the range of 293 o 573k ii. slip geometry and
  structure of slip bands.
\newblock {\em Physica Status Solidi B}, 46:203--215, 1971.

\bibitem[RVC{\etalchar{+}}17]{Rodney:2017}
D.~Rodney, L.~Ventelon, E.~Clouet, L.~Pizzagalli, and F.~Willaime.
\newblock Ab initio modeling of dislocation core properties in metals and
  semiconductors.
\newblock {\em Acta Materialia}, 124:663--659, 2017.

\bibitem[RZH{\etalchar{+}}98]{Rhee:1998}
M.~Rhee, H.~M. Zbib, J.~P. Hirth, H.~Huang, and T.~de~la Rubia.
\newblock Models for long-/short-range interactions and cross slip in 3d
  dislocation simulation of bcc single crystals.
\newblock {\em Modelling and Simulation in Materials Science and Engineering},
  6:467--492, 1998.

\bibitem[SBA12]{Stukowski2012}
Alexander Stukowski, Vasily~V Bulatov, and Athanasios Arsenlis.
\newblock {Automated identification and indexing of dislocations in crystal
  interfaces}.
\newblock {\em Modelling and Simulation in Materials Science and Engineering},
  20(8):085007, 2012.

\bibitem[Sch05]{Schoeck:2005}
G.~Schoeck.
\newblock The peierls model: Progress and limitations.
\newblock {\em Materials Science and Engineering A}, 400-401:7--17, 2005.

\bibitem[SGGM75]{Shields:1975}
J.~A. Shields, S.~H. Goods, R.~Gibala, and T.~E. Mitchell.
\newblock Deformation of high purity tantalum single crystals at 4.2k.
\newblock {\em Materials Science and Engineering}, 20:71--81, 1975.

\bibitem[SLB{\etalchar{+}}07]{Shehadeh:2007}
M.~A. Shehadeh, G.~Lu, S.~Banerjee, N.~Kioussis, and N.~Ghoniem.
\newblock Dislocation transmission across the cu/ni interface: a hybrid
  atomistic-continuum study.
\newblock {\em Philosophical Magazine}, 87(10):1513--1529, 2007.

\bibitem[Ste09]{Steinbach:2009}
I.~Steinbach.
\newblock Phase-field models in materials science.
\newblock {\em Modelling and Simulation in Materials Science and Engineering},
  17:073001, 2009.

\bibitem[Stu09]{Stukowski2009}
Alexander Stukowski.
\newblock {Visualization and analysis of atomistic simulation data with
  OVITO–the Open Visualization Tool}.
\newblock {\em Modelling and Simulation in Materials Science and Engineering},
  18(1):015012--1--7; Also see http://www.ovito.org/, 2009.

\bibitem[Stu14]{Stukowski2014}
Alexander Stukowski.
\newblock {A triangulation-based method to identify dislocations in atomistic
  models}.
\newblock {\em Journal of the Mechanics and Physics of Solids}, 70(1):314--319,
  2014.

\bibitem[TKC98]{Tang:1998}
M.~Tang, L.~P. Kubin, and G.~R. Canova.
\newblock Dislocation mobility and the mechanical response of bcc single
  crystals: A mesoscopic approach.
\newblock {\em Acta Materialia}, 46(9):3221--3235, 1998.

\bibitem[UW75]{Urabe:1975}
N.~Urabe and J.~Weertman.
\newblock Dislocation mobility in potassium and iron single crystals.
\newblock {\em Materials Science and Engineering}, 18:41--49, 1975.

\bibitem[Vit68]{Vitek:1968}
V.~Vitek.
\newblock Intrinsic stacking faults in body-centered cubic crystals.
\newblock {\em Philosophical Magazine}, 18(154):773--786, 1968.

\bibitem[Vit74]{Vitek:1974}
V.~Vitek.
\newblock Theory of core structures of dislocations in bcc metals.
\newblock {\em Cryst. Latt. Defects}, 5:1, 1974.

\bibitem[VP08]{Vitek:2008}
V.~Vitek and V.~Paidar.
\newblock Non-planar dislocation cores: A ubiquitous phenomenon affecting
  mechanical properties of crystalline materials.
\newblock In J.~P. Hirth, editor, {\em Dislocations in Solids}, chapter~87,
  pages 439--514. Elsevier, 2008.

\bibitem[WB11]{ZWang:2011}
Z.~Q. Wang and I.~J. Beyerlein.
\newblock An atomistically-informed dislocation dynamics model for the plastic
  anisotropy and tension-compression asymmetry of bcc metals.
\newblock {\em International Journal of Plasticity}, 27(10):1471--1484, 2011.

\bibitem[WBB13]{Weinberger2:2013}
C.~R. Weinberger, B.~L. Boyce, and C.~C. Battaile.
\newblock Slip planes in bcc transition metals.
\newblock {\em International Materials Reviews}, 58(5):296--314, 2013.

\bibitem[WBL07]{ZWang2:2007}
Z.~Q. Wang, I.~J. Beyerlein, and R.~LeSar.
\newblock The importance of cross-slip in high-rate deformation.
\newblock {\em Modelling and Simulation in Materials Science and Engineering},
  15(6):675--690, 2007.

\bibitem[WBL08]{ZWang:2008}
Z.~Q. Wang, I.~J. Beyerlein, and R.~LeSar.
\newblock Slip band formation and mobile dislocation density generation in high
  rate deformation of single fcc crystals.
\newblock {\em Philosophical Magazine}, 88:1321--1343, 2008.

\bibitem[WBL09]{ZWang:2009}
Z.~Q. Wang, I.~J. Beyerlein, and R.~LeSar.
\newblock Plastic anisotropy in fcc single crystals in high rate deformation.
\newblock {\em International Journal of Plasticity}, 25:26--48, 2009.

\bibitem[WBT14]{JWang:2014}
J.~Wang, I.~J. Beyerlein, and C.~N. Tom\'{e}.
\newblock Reactions of lattice dislocations with grain boundaries in mg:
  implications on the micro scale from atomic-scale.
\newblock {\em International Journal of Plasticity}, 56:156--172, 2014.

\bibitem[WFVdGN02]{Weygand:2002}
D.~Weygand, L.~H. Friedman, E.~Van~der Giessen, and A.~Needleman.
\newblock { Aspects of boundary-value problem solutions with three-dimensional
  dislocation dynamics}.
\newblock {\em Modelling and Simulation in Materials Science and Engineering},
  10:437, 2002.

\bibitem[WJCK01]{Wang2001}
Y.~U. Wang, Y.~M. Jin, A.~M. Cuiti{\~{n}}o, and A.~G. Khachaturyan.
\newblock {Nanoscale phase field microelasticity theory of dislocations: Model
  and 3D simulations}.
\newblock {\em Acta Materialia}, 49(10):1847--1857, 2001.

\bibitem[WScGI04]{GWang:2004}
G.~Wang, A.~Strachan, T.~\c{C}a\u{g}in, and W.~A. Goddard~III.
\newblock Calculating the peierls energy and peierls stress from atomistic
  simulations of screw dislocation dynamics: application to bcc tantalum.
\newblock {\em Modelling and Simulation in Materials Science and Engineering},
  12:S371--S389, 2004.

\bibitem[WTF13]{Weinberger:2013}
C.~R. Weinberger, G.~J. Tucker, and S.~M. Foiles.
\newblock Peierls potential of screw dislocations in bcc transition metals:
  Predictions from density functional theory.
\newblock {\em Physical Review B}, 87:054114, 2013.

\bibitem[YLH13]{Yang:2013}
H.~Yang, Z.~Li, and M.~Huang.
\newblock Modeling dislocation cutting the precipitate in nickel-based single
  crystal superalloy via the discrete dislocation dynamics with sisf
  dissociation scheme.
\newblock {\em Computational Materials Science}, 75:52--59, 2013.

\bibitem[ZBGW11]{Zhang2:2011}
R.~F. Zhang, I.~J. Beyerlein, T.~C. Germann, and J.~Wang.
\newblock Deformation twinning in bcc metals under shock loading: a challenge
  to empirical potentials.
\newblock {\em Philosophical Magazine Letters}, 91(12):731--740, 2011.

\bibitem[ZBL11]{Zhou:2011}
C.~Zhou, I.~J. Beyerlein, and R.~LeSar.
\newblock Plastic deformation mechanisms of fcc single crystals at small
  scales.
\newblock {\em Acta Materialia}, 59(20):7673--7682, 2011.

\bibitem[ZCK19]{Zeng:2019}
Y.~Zeng, X.~Cai, and M.~Koslowski.
\newblock Effects of the stacking fault energy fluctuations on the
  strengthening of alloys.
\newblock {\em Acta Materialia}, 164:1--11, 2019.

\bibitem[ZdlRRH00]{Zbib:2000}
H.~M. Zbib, T.~D. de~la Rubia, M.~Rhee, and J.~P. Hirth.
\newblock 3d dislocation dynamics: Stress strain behavior and hardening
  mechanisms in fcc and bcc metals.
\newblock {\em Journal of Nuclear Materials}, 276:154--165, 2000.

\bibitem[ZHBK16]{Zeng:2016}
Y.~Zeng, A.~Hunter, I.~J. Beyerlein, and M.~Koslowski.
\newblock A phase field dislocation dynamics model for a bicrystal interface
  system: An investigation into dislocation slip transmission across
  cube-on-cube interfaces.
\newblock {\em International Journal of Plasticity}, 79:293--313, 2016.

\bibitem[ZOAB11]{Zbib:2011}
H.~M. Zbib, C.~T. Overman, F.~Akasheh, and D.~Bahr.
\newblock Analysis of plastic deformation in nanoscale metallic multilayers
  with coherent and incoherent interfaces.
\newblock {\em International Journal of Plasticity}, 27(10):1618--1639, 2011.

\bibitem[ZRSOB17]{Zepeda-Ruiz2017}
Luis~A. Zepeda-Ruiz, Alexander Stukowski, Tomas Oppelstrup, and Vasily~V.
  Bulatov.
\newblock {Probing the limits of metal plasticity with molecular dynamics
  simulations}.
\newblock {\em Nature}, 550(7677):492--495, 2017.

\end{thebibliography}

\end{document}